\documentclass{article}
\usepackage[utf8]{inputenc}  
\usepackage[T1]{fontenc}
\usepackage{hyperref}        
\usepackage[super,sort&compress]{natbib} 
\usepackage{geometry}        
\usepackage{tabularx}
\usepackage{graphicx}
\usepackage{xcolor}          
\newif\ifshowchanges
\showchangesfalse
\definecolor{revcolor}{RGB}{139,0,0} 
\newcommand{\rev}[1]{\ifshowchanges{\color{revcolor}#1}\else#1\fi}

\title{Beyond Structure: Revolutionising Materials Discovery via AI-Driven Synthesis Protocol-Property Relationships}
\author{Guillaume Lambard\thanks{Corresponding author. Email: \href{mailto:LAMBARD.Guillaume@nims.go.jp}{LAMBARD.Guillaume@nims.go.jp}}\\
\small Data-driven materials design group, Center for Basic Research on Materials (CBRM)\\
\small National Institute for Materials Science (NIMS), Namiki 1-1, Tsukuba, Ibaraki, 305-0044, Japan}
\date{\today}

\begin{document}
\maketitle
%
\begin{abstract}
The current structure-centric paradigm in artificial intelligence (AI)-driven materials discovery, despite delivering thousands of candidate structures, is stalling at a critical barrier: the synthesizability gap. We argue that closing this gap demands a pivot to a synthesis-first paradigm in which executable synthesis protocols, not just atomic configurations, are treated as primary design variables. We outline a roadmap built on three pillars: (i) representing synthesis procedures as machine-readable protocols, (ii) deploying generative and inverse-design models to propose actionable reaction pathways and recipes, and (iii) integrating closed-loop optimisation to refine protocols against experimental realities and sustainability constraints. Framed in terms of the causal backbone $P\!\to\!X\!\to\!y$ from protocol $P$ to structure $X$ and properties $y$, this perspective sets out methodological building blocks, standards needs and self-driving laboratory (SDL) integration strategies to accelerate reproducible, data-first materials discovery.
\end{abstract}

\section{Introduction}

Advanced materials underpin solutions to pressing global challenges in sustainable energy, healthcare, environmental remediation and information technologies, and progress increasingly depends on discovering materials with targeted properties faster than conventional trial-and-error allows.~\cite{ref1} For decades, computational materials science has been dominated by a structure-property paradigm grounded in the premise that \emph{atomic structure dictates observable properties}. High-throughput density functional theory (DFT) workflows and large databases such as the Materials Project, AFLOW and the Open Quantum Materials Database (OQMD) have delivered thousands of low-enthalpy candidates and enabled remarkable computational advances, yet many theoretically promising structures remain unrealised in the laboratory.~\cite{refMGI,ref6,refAFLOW,refOQMD}

The field now confronts a decisive bottleneck: structure-centric approaches—whether traditional DFT screening or modern AI generative models often trained on DFT data—routinely falter when their predictions meet experimental synthesis, exposing a critical \emph{synthesizability gap}. Fewer than 2\,\% of over 50\,000 low-enthalpy phases identified in high-throughput surveys have been realised experimentally, underscoring the limitations of pipelines that neglect kinetics, precursor availability and practical constraints.~\cite{refStatGap} We contend that closing this gap demands a synthesis-first paradigm in which executable synthesis protocols, not just atomic configurations, are treated as the primary design variables.

This motivates a shift toward a \emph{synthesis protocol-property} framework centred on the causal backbone $P\!\to\!X\!\to\!y$, where a synthesis protocol $P$ maps to structure, phase or morphology $X$ and ultimately to properties $y$. For clarity, we define $P$ as the complete, machine-readable specification of a material recipe—precursors, stoichiometries, sequence of operations (\textsc{ADD}, \textsc{HEAT}, \textsc{COOL}, \textsc{FILTER}, \dots), and quantitative conditions (temperature, time, atmosphere, pressure, pH). Whether a candidate originates from a deep generative model or an \emph{ab~initio} evolutionary search, thermodynamic stability alone is an insufficient metric of practical viability; hybrid pipelines that couple structure-centric tools with protocol-aware planners are needed to bridge from virtual candidates to executable recipes. Related ideas already exist in autonomous experimentation, retrosynthesis planning and closed-loop reaction optimisation.~\cite{ref20,refASKCOS,refColeyReview,refMinerva} Our claim is not that these ingredients are themselves new, but that in inorganic materials discovery they remain insufficiently unified by a protocol-first formalism in which synthesis procedures are treated as primary design variables and linked explicitly to intermediate structure through $P\!\to\!X\!\to\!y$. While the concept of protocol-centric design is universal, implementation differs between organic molecular synthesis and inorganic/solid-state materials. This perspective targets the latter, using organic tools (e.g., the Simplified Molecular Input Line Entry System (SMILES), SELF-referencIng Embedded Strings (SELFIES), and retrosynthesis planners) as analogies where instructive but not assuming their direct transfer to inorganic synthesis governed by phase equilibria, non-equilibrium processing and transport-limited kinetics.

To realise this vision, we pursue three specific objectives: (i) to critically assess the limitations of current structure-centric 
generative and predictive AI models; (ii) to formalise the synthesis protocol-property framework and survey the representations and learning algorithms that enable it; and (iii) to chart open challenges while outlining research directions toward a fully autonomous, synthesis-aware materials discovery ecosystem. In this perspective, we outline the AI methodologies—from protocol representation and generative synthesis planning to closed-loop optimisation—that will drive this synthesis-first paradigm and envision a future where AI seamlessly designs complete, experimentally realisable material recipes.

This perspective centres \emph{machine-readable synthesis protocols} as first-class design objects, detailing representations, inverse-design methods and closed-loop strategies that directly support reproducible, data-first discovery in automated and human-in-the-loop labs.

\rev{
\noindent\textbf{Novel contributions of this perspective.}
Although many materials scientists recognise that synthesis governs what is experimentally accessible, we argue that the \emph{operational} consequences for AI workflows remain under-specified. This perspective therefore emphasises several concrete contributions:
\begin{itemize}
  \item \textbf{A protocol-first formalism anchored in \(P\!\to\!X\!\to\!y\):} we treat protocols as first-class design objects and use \(P\!\to\!X\!\to\!y\) to separate prediction-only workflows from mechanistic, characterisation-aware modelling.
  \item \textbf{A systems view spanning models, automation and interoperability:} we connect representations and inverse design to the practical realities of self-driving laboratories, including the need for interoperable protocol execution and provenance-aware logging across platforms.
  \item \textbf{Inorganic-specific constraints and failure modes:} we highlight modelling and data issues that are distinctive for inorganic/solid-state synthesis (multiphase outcomes, path dependence, reactor effects, and sparse or biased protocol corpora), and show how these shape representation and learning choices.
\end{itemize}
More broadly, synthesis- and process-centred representations also resemble how human experimentalists plan, troubleshoot and refine procedures, complementing structure-centric surrogates that often approximate upstream simulations.
\rev{This framing differs from treating autonomous laboratories merely as execution layers or treating retrosynthesis tools merely as route generators: here, representation, forward modelling, inverse design, characterisation and execution are posed as one coupled protocol-design problem.}
}

These objectives motivate the structure of the present perspective. Building on the challenges above, Section~\ref{sec:generative} assesses limitations of structure-centric generative AI. Section~\ref{sec:synthesis} lays out the synthesis-centric paradigm. Section~\ref{sec:methods} surveys enabling artificial intelligence and machine learning (AI/ML) methodologies, while Section~\ref{sec:cases} provides application case studies. Section~\ref{sec:challenges} discusses outstanding challenges, and Section~\ref{sec:conclusion} concludes with a future outlook.

\begin{figure}[ht]
  \centering
  \includegraphics[width=0.9\textwidth]{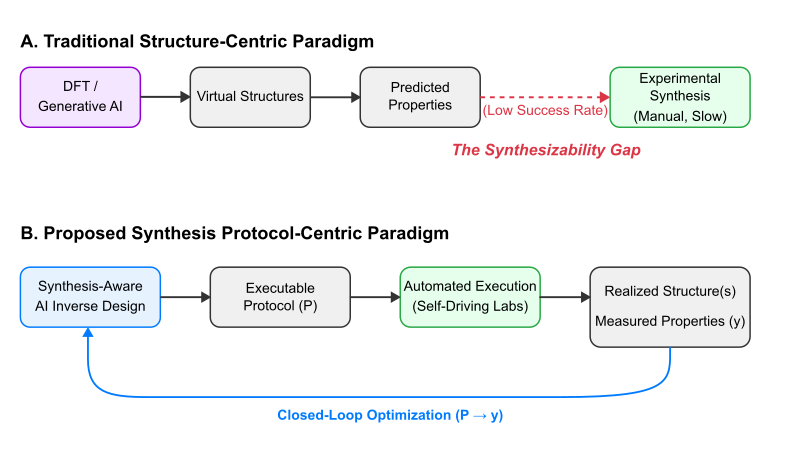}
  \caption{Paradigm shift from structure-centric to synthesis protocol-centric discovery. (A) Conventional workflow emphasises virtual structure generation and property prediction, leaving a dashed \emph{synthesizability gap} to experimental realisation. (B) The proposed workflow elevates executable synthesis protocols to first-class design objects and closes the loop via autonomous experimentation.}
  \label{fig:paradigm}
\end{figure}


\section{Generative AI for Materials Discovery: The Structure-Property Paradigm and Its Limitations}
\label{sec:generative}

Having established the synthesizability gap as a fundamental limitation of structure-centric approaches, we now examine the current state of generative AI for materials discovery. This critical assessment reveals both the remarkable achievements and inherent constraints of structure-property models, providing the foundation for our synthesis-first alternative. While organic retrosynthesis heuristics (e.g., synthetic accessibility, synthetic complexity, and retrosynthetic accessibility scores) and planners (e.g., ASKCOS and AiZynthFinder~\cite{refASKCOS,refAiZynth}) are mature for molecular synthesis, they are only partially applicable to solid-state and solvothermal materials. In this section we focus on inorganic-relevant approaches and treat organic tools as instructive analogues rather than direct solutions.

\subsection{Overview of Generative Models for Materials Structures}

Generative AI models have emerged as powerful tools for exploring chemical space and proposing novel material structures \emph{de~novo}.\cite{ref16} 
These models learn underlying patterns and distributions from existing materials data and generate new instances with potentially desirable properties. For inorganic crystals, symmetry-aware, crystallographic information file (CIF)-based and graph representations (e.g., crystal graph convolution networks and equivariant architectures) are commonly used alongside composition-based encodings.
Several architectures are widely adopted:

\begin{itemize}
    \item \textbf{Variational Autoencoders (VAEs):} VAEs learn a compressed, continuous latent representation of input structures 
    (e.g., molecular graphs, crystal structures). By sampling points from this latent space and decoding them, novel structures can be generated. 
    VAEs have been applied to design molecules, polymers, and even porous materials such as metal-organic frameworks (MOFs).~\cite{ref7}
    \item \textbf{Generative Adversarial Networks (GANs):} GANs employ a two-player game framework. A generator network creates candidate structures, 
    while a discriminator network attempts to distinguish these generated structures from real ones in the training data. Through adversarial training, 
    the generator learns to produce increasingly realistic and diverse structures.~\cite{ref19}
    \item \textbf{Diffusion Models:} Inspired by non-equilibrium thermodynamics, diffusion models learn to reverse a process that gradually adds noise 
    to data until only noise remains. By starting with noise and applying the learned reverse process, highly realistic and diverse structures can be 
    generated. These models currently achieve state-of-the-art performance for molecular generation.~\cite{ref16}
    \item \textbf{Autoregressive Models:} These models generate structures sequentially, predicting the next atom, bond, or fragment based on the 
    previously generated parts. Recurrent neural networks and Transformer architectures operating on string representations such as SMILES or SELFIES 
    belong in this class.~\cite{ref16}
    \item \textbf{Flow-based Models:} These learn an explicit, invertible transformation between the complex data distribution and a simple base 
    distribution (e.g., a standard Gaussian). Generation involves sampling from the base distribution and applying the inverse transformation.~\cite{ref16}
\end{itemize}

Beyond these archetypes, recent studies report strong performance from energy-based models, Transformer decoders, and 3D equivariant architectures, extending generative design capabilities.~\cite{refHoogeboom,refEGNN}

These models operate on different structural representations. For molecules and polymers, SMILES, SELFIES and graph representations are common; for 
inorganic crystals, CIF or graph-based abstractions are typical.  Such generative approaches have successfully 
proposed novel candidates across drug discovery and energy materials.

\subsection{The Achilles' Heel: The Synthesizability Gap}

Despite their success in generating structurally novel and potentially high-performing materials \emph{in~silico}, the practical utility of 
structure-centric generative models is limited by the synthesizability gap.~\cite{ref2} Experimental chemists frequently struggle to realise 
these computationally designed structures in the laboratory.  Key contributing factors include: 

\begin{itemize}
    \item \emph{Unknown or infeasible reaction pathways}: No known chemical transformation may exist that converts available starting materials into 
    the target structure.  Generative models often output molecules that violate established reactivity principles.~\cite{refGao}
    \item \emph{Precursor availability and cost}: Required building blocks might be unavailable, prohibitively expensive, or unstable.~\cite{ref45}
    \item \emph{Impractical reaction conditions}: Predicted syntheses may require extreme temperatures, pressures, or specialised equipment.
    \item \emph{Kinetic traps and competing reactions}: Models optimising for thermodynamic stability neglect kinetics; desired products can be 
    kinetically inaccessible or obscured by side reactions.~\cite{ref28}
    \item \emph{Purification and isolation challenges}: Even if formed, the target may be difficult to separate from complex mixtures.
    \item \emph{Scalability}: A route that succeeds on milligram scale may fail at scale-up.
    \item \emph{Regulatory, safety or environmental constraints}: some suggested precursors are toxic, explosive or legally restricted, barring 
    practical implementation.~\cite{refEPA}
\end{itemize}

Thus, structure-only optimisation implicitly assumes synthesis is a solvable downstream task, whereas in reality synthetic constraints define the 
accessible chemical space. Even advanced surrogate models—such as energy-conserving equivariant graph neural networks (\textsc{MACE}~\cite{refMACE}) that provide symmetry-respecting predictions with near-DFT accuracy—remain fundamentally limited by their structure-centric foundation.

\subsection{Critiquing Structure-Centric AI}

To compensate, post-hoc heuristics such as the synthetic accessibility score (\textsc{SA}Score)~\cite{ref45}, synthetic complexity score (\textsc{SC}Score)~\cite{refSCScore}, and retrosynthetic accessibility score (\textsc{RA}Score)~\cite{refRAscore} are often applied to filter generative outputs. These heuristics, however, are only loosely correlated
with true experimental difficulty (e.g., $r\approx0.3$ for \textsc{SA}Score vs.~expert labels~\cite{refGao}). Moreover, although graphics processing unit (GPU)-accelerated 
Monte-Carlo planners such as \textsc{AiZynthFinder}~\cite{refAiZynth} can now process $\mathcal{O}(10^{4})$ molecules per hour on a single node, 
explicit retrosynthesis for millions of candidates remains prohibitive at library-generation scale.

\rev{A complementary strategy is to predict synthesizability directly with supervised learning, treating it as a classification or ranking problem rather than as a downstream constraint applied post hoc. For crystalline materials, deep learning models trained on databases of known compounds have been used to estimate whether hypothetical compositions or structures are likely to be experimentally realisable, providing a fast prior that can be combined with generative design or DFT screening.~\cite{refSynthDL} Such predictors are promising but must be interpreted carefully: they inherit biases in what the community has attempted and reported, and the target label ("synthesizable") is itself time-, platform- and effort-dependent.}

Ultimately, the synthesizability gap reveals a fundamental limitation of the structure-property paradigm in \emph{de~novo} design: it explores 
theoretical chemical space without embedding the constraints that bound experimental reality.

Table~\ref{tab:generative} summarises major generative model families for structure design, highlighting their strengths and synthesizability limitations.

Therefore, while generative AI has achieved remarkable success in proposing novel structures with desirable computed properties, the persistent synthesizability gap exposes an architectural flaw in the structure-first approach. Post-hoc synthesizability filters provide only weak correlation with experimental difficulty, and computational retrosynthesis remains prohibitively expensive at scale. This systematic failure across all major generative architectures—from VAEs to diffusion models—indicates that the solution lies not in incremental improvements to structure-centric methods, but in a fundamental paradigm shift that treats synthesis protocols as primary design objects. The following section outlines this synthesis-first framework and its potential to bridge the gap between computational prediction and laboratory reality.

\begin{table}[ht]
\centering
\setlength{\tabcolsep}{3pt}           
\begin{tabularx}{\textwidth}{>{\raggedright\arraybackslash}
                             >{\raggedright\arraybackslash}p{2.5cm}
                             >{\raggedright\arraybackslash}p{2.8cm}
                             >{\raggedright\arraybackslash}p{1.7cm}
                             >{\raggedright\arraybackslash}X
                             >{\raggedright\arraybackslash}X}
\hline
\textbf{Model type} & \textbf{Input repr. (molecular/inorganic)} & \textbf{Output} &
\textbf{Strengths} & \textbf{Limitations (synthesizability)}\\
\hline
VAE & Graphs, SMILES / CIF, symmetry-aware graphs & Structure & Smooth latent space, property optimisation & May generate unsynthesizable structures; mode collapse \\
GAN & Latent vector, graphs / crystal graphs & Structure & Highly realistic, diverse outputs & Training instability; mode collapse; synthetic tractability unclear \\
Diffusion & Graphs, point clouds / crystal lattices & Structure & High quality and diversity & Expensive training/sampling (\(\sim10^{3}\) steps); no synth. guarantee \\
Autoregressive & SMILES, SELFIES / composition-sequence & Structure & Effective sequential generation & Sensitive to representation; validity issues \\
Flow & Graphs, SMILES / symmetry-aware graphs & Structure & Exact likelihood; invertible & Complex for discrete data; synth. not explicit$^{\dagger}$ \\
\hline
\multicolumn{5}{l}{\footnotesize{$^{\dagger}$Grammar-constrained flow models partially alleviate this limitation.\cite{refJacques}}} \\
\end{tabularx}
\caption{Generative model families used for materials structure design together with their principal strengths and synthesizability limitations.}
\label{tab:generative}
\end{table}
 
Additional protocol representations, including domain-specific language (DSL) and ontology formats as well as multimodal embeddings, are discussed in Section~\ref{sec:methods}.
 
 
 \section{The Synthesis Protocol--Property Paradigm: A Necessary Shift}
 \label{sec:synthesis}
 
 The persistent failures of structure-centric approaches demand a fundamental paradigm shift. We propose elevating synthesis protocols to first-class design objects in a synthesis protocol--property framework that embeds experimental feasibility from the outset. This approach directly addresses the synthesizability gap by treating the recipe—not just the product—as the primary design variable.
 
While one might object that \emph{structure is fundamental} (atomic arrangement ultimately determines material properties), our paradigm treats structure as an indispensable intermediate outcome produced by a synthesis protocol, rather than an abstract starting point divorced from lab realities. Similarly, critics may argue that \emph{synthesis data doesn't exist} at the necessary scale. While data scarcity poses challenges (see Section~\ref{sec:challenges}), advances in self-driving laboratories and natural language processing (NLP)-driven literature mining are rapidly generating rich protocol-property datasets. This confluence makes the synthesis-first shift both timely and feasible.
 
 \subsection{Conceptual Framework}
 
 In the synthesis-centric view, a material's structure is treated as an \emph{intermediate outcome} that emerges from executing a specific synthesis protocol. Characterising this intermediate—e.g., through \emph{ex situ} diffraction or \emph{in~situ} spectroscopy—remains scientifically vital, because the causal chain we ultimately seek to model is $P \rightarrow X \rightarrow y$, where $X$ denotes structure, phase, or morphology and $y$ the resulting properties. 
 
\rev{Two operational modes are common in practice: (i) a purely predictive mode that prioritises $P\!\to\!y$ to obtain target properties efficiently, and (ii) a characterisation-aware mode that uses $P\!\to\!X\!\to\!y$ to learn \emph{why} a protocol yields a given outcome. Direct $P\!\to\!y$ models can succeed when the relevant intermediate structure is either weakly varying or implicitly absorbed into the training distribution, but explicit treatment of $X$ becomes important when phase selection, morphology evolution, defect formation or multi-step transformations mediate the final property. While additional $X$ data can in principle improve predictive performance, integrating heterogeneous and often sparse characterisation into $P\!\to\!y$ models is non-trivial; in many settings the primary benefit is mechanistic understanding and theory-building.}
 
Rich structural data therefore act as mechanistic ground truth that fuels generalisable models. The core ML tasks therefore become (i)~the forward mapping $P \;\rightarrow\; y$ to predict properties $y$ from a protocol and (ii)~the inverse mapping $y^{\star} \;\rightarrow\; P^{\star}$ to design protocols $P^{\star}$ that deliver target properties $y^{\star}$. Because protocols explicitly encode precursors, sequences, temperatures, times, catalysts and post-treatments, the resulting models are automatically grounded in experimental reality. In practice a single protocol may yield \emph{multiple} polymorphs or morphologies, so probabilistic forward models that capture this multimodality, together with controllers that repeatedly infer structure-sensitive proxies of $X$, remain an active research frontier.

\noindent\textbf{Task taxonomy: optimisation vs. de~novo protocol design.}
\rev{We distinguish two tasks that are often conflated: (i) \emph{process/parameter optimisation} for a \emph{known} material and fixed procedural scaffold (continuous and discrete variables such as temperature, time, concentrations), and (ii) \emph{de~novo protocol design}, i.e., constructing the \emph{sequence of operations}, reagent choices, and intermediate targets for a \emph{new} material. Task (i) is well-suited to Bayesian optimisation (BO) in a constrained design space; Task (ii) is a high-dimensional, sequential decision problem that remains nascent for inorganic materials. We use this taxonomy throughout the rest of the perspective (see Section~\ref{sec:cases}).}

\noindent\textbf{Characterisation as the rate-limiting step.}
In practice, estimating $X$ (phase, microstructure, defects, morphology) is often the slowest and most manual component of the loop. Building robust $P\!\to\!X\!\to\!y$ models therefore hinges on advances in high-throughput \emph{ex~situ} pipelines (automated Rietveld refinement, computer vision for morphology) and \emph{in~situ/operando} probes (synchrotron diffraction, small-angle and wide-angle X-ray scattering (SAXS/WAXS), inline spectroscopy, inline electron microscopy). These modalities provide time-resolved constraints that reduce ambiguity in $X$, enable causal attribution, and materially accelerate closed-loop optimisation.

\noindent\textbf{On the intrinsic difficulty of $P\!\to\!X$.}
Predicting $X$ from $P$—spanning nucleation barriers, multi-step reaction pathways, polymorphic transformations, grain growth and porosity evolution, non-stoichiometry, and diffusion-limited transport—is frequently \emph{harder} than the traditional structure$\,\to\,$property task. The mapping is path-dependent and governed by non-equilibrium kinetics, multi-scale transport, and reactor/vessel effects. Practical strategies therefore require: (i) rich, time-resolved ground truth via \emph{in~situ/operando} probes to disambiguate pathways; (ii) hybrid, physics-guided machine learning with simulators acting as priors or constraints; (iii) multi-fidelity learning across literature, simulator outputs, and self-driving laboratory (SDL) data; and (iv) explicit uncertainty quantification to guide information gain. \rev{Validation must likewise go beyond random train/test splits and include chemically held-out systems, cross-laboratory transfer, calibration against operando or \emph{ex~situ} characterisation, and ablations that compare direct $P\!\to\!y$ predictors with variants that explicitly incorporate $X$.}
 
 \subsection{Advantages of the Synthesis-Centric Approach}

Within the synthesis protocol--property framework, several advantages over traditional structure-centric approaches emerge:

 \begin{itemize}
    \item \textbf{Experimental executability, not guaranteed synthesizability}: Protocols generated within a constrained action/reaction grammar are, in principle and subject to platform and safety constraints, \emph{executable} in a laboratory setting. Executability does not imply the target will be synthesised; rather, it confines search to experimentally actionable procedures, improving relevance and enabling efficient optimisation over real operations.~\cite{refXDL} \rev{In practice, higher success probability comes not from executability alone but from combining such constraints with uncertainty-aware model selection, structure-sensitive feedback, and iterative closed-loop refinement.}
     \item \textbf{Process parameters included}: Properties that depend sensitively on temperature ramps, solvent, pH or ageing time can be 
     learned because these variables are explicit inputs.~\cite{ref70}
     \item \textbf{Interface with automation}: Protocol representations map naturally to robot instructions, enabling closed-loop, SDLs.~\cite{refMinerva}
     \item \textbf{True inverse design}: Instead of suggesting exotic structures and leaving synthesis an open problem, the model outputs a 
     \emph{recipe} that can be run the same day.~\cite{refColeyReview}
     \item \textbf{Deeper scientific insight}: Correlating process variables with performance uncovers non-equilibrium effects, defect formation 
     and morphology control invisible to equilibrium structure-only models.~\cite{refEngel}
    \item \textbf{Sustainability and economics}: Green-chemistry metrics (E-factor, process mass intensity (PMI), energy footprint) and economic costs (precursor cost, throughput, labour, equipment amortisation) can be included directly in multi-objective optimisation, steering design toward practically deployable solutions.~\cite{refGreenMetrics}
 \end{itemize}
 
 These conceptual advantages position the synthesis-centric paradigm as a direct solution to the synthesizability gap. By embedding experimental constraints from the outset, this approach promises to bridge the persistent divide between computational prediction and laboratory reality. The following section details the AI/ML methodologies that enable this paradigm shift.
  
 
 \section{Enabling AI/ML Methodologies for Synthesis-Driven Discovery}
 \label{sec:methods}
 
 Realising the vision of a synthesis-first paradigm requires a new class of AI/ML tools. The emerging toolbox, capable of \emph{understanding}, generating, and optimising procedural synthesis data, is already taking shape. Three key technical challenges emerge: (i)~how to represent a synthesis protocol, (ii)~how to predict properties from that representation, and (iii)~how to invert the model to design new protocols.
 
 \subsection{Representing Synthesis Protocols}
 
Common synthesis protocol representations currently in literature span four main paradigms, each reflecting distinct trade-offs between expressivity and structure. Text- and NLP-based formats treat the protocol as raw procedural prose or reaction SMILES strings, enabling transformer models pretrained on large chemical corpora to learn directly from unstructured text. Graph-based representations abstract protocols as reaction graphs—where molecules, operations, and vessels form nodes connected by temporal or chemical relationships—making them ideal for graph neural networks (GNNs). Action-sequence encodings view protocols as chronological lists of primitive operations (e.g., \textsc{ADD}, \textsc{HEAT}, \textsc{FILTER}), aligning naturally with reinforcement learning (RL) frameworks and robotic execution. Finally, tabular or vector representations distil well-defined design-of-experiments variables—such as temperature, time, and pH—into continuous vectors, facilitating efficient optimisation via conventional ML or BO.

Beyond these four canonical families, several specialised representation strategies are particularly relevant:

\begin{itemize}
  \item \textbf{DSL and ontology representations:} Machine-executable domain-specific languages and standards such as the chemical description language (XDL), Autoprotocol, Protocol Activity Modeling Language (PAML) and Standardization in Laboratory Automation (SiLA)~2 provide structured grammars that map directly onto robotic hardware while maintaining human readability, while Analytical Information Markup Language (AnIML) supports standardised analytical data exchange.~\cite{refXDL,refAutoprotocol,refPAML,refSiLA2,refAnIML} Knowledge-graph approaches leveraging domain ontologies such as the Reaction Ontology (RXNO) and Chemical Entities of Biological Interest (ChEBI) likewise enable symbolic reasoning and constraint checking across large corpora.~\cite{refRXNO,refChEBI}
  \item \textbf{Multimodal embeddings:} Emerging work fuses free-text protocols with time-resolved sensor streams (spectra, images) to create rich joint embeddings that support closed-loop optimisation and anomaly detection.~\cite{refMacLeod}
  \item \textbf{Inorganic-specific modelling considerations:} Unlike molecular retrosynthesis, inorganic synthesis often lacks discrete reaction rules and features continuous non-stoichiometry, high-temperature processing, and diffusion-limited kinetics. Effective encodings therefore incorporate (i) thermochemical and phase-field-inspired features (activities, phase stability margins, oxygen chemical potential), (ii) transport-aware parameters (thermal ramps, dwell times, gas/flow conditions), (iii) reactor and furnace geometry (crucible materials, atmosphere zones, loading configuration), and (iv) tolerance for continuous composition spaces. Graph representations should include vessels, solids, melts, atmospheres and contact interfaces; action-sequence grammars should encode temperature/pressure profiles, grinding/milling, pelletising, soaking and controlled cooling. These additions centre solid-state realities (diffusion, sintering, volatilisation) often absent from molecular encodings.
\end{itemize}

\rev{Taken together, no single representation is universally best for inorganic synthesis. For many solid-state or thin-film optimisation loops, a hybrid of action sequences and tabular process variables is the most pragmatic starting point because it preserves ordered operations while exposing the thermochemical controls most often tuned in practice. Graph or DSL-based formats become especially valuable when vessel context, equipment logic or cross-platform execution constraints must be preserved explicitly, whereas text-centric representations remain indispensable for mining legacy literature corpora.}
 
\subsection{Predicting Properties from Protocols}
 
Once protocols are encoded, supervised models can be trained. Common choices are GNNs for graph inputs, Transformers/long short-term memory (LSTM) networks for sequences, and gradient-boosted trees or feed-forward networks for tabular data. Model accuracy is limited less by algorithmic nuance than by the scarcity and quality of paired protocol-property datasets—an area where SDLs can contribute enormously.

Beyond model architectures, reproducibility is equally critical. Protocol-aware models should be accompanied by accessible code and data: wherever possible, authors should deposit protocol corpora, model weights, training scripts and figure-generation notebooks in open repositories such as Zenodo or institutional archives, providing digital object identifiers (DOIs) that can be cited alongside this perspective. Even when no new experimental datasets are generated, releasing synthetic benchmark corpora, configuration files and source data for figures materially lowers the barrier to reuse and supports the data-review processes now expected by journals.
 
\rev{Physics-informed neural networks (PINNs) and other hybrid surrogate models that embed differential-equation constraints offer a promising but challenging avenue for synthesis modelling; while they are well developed in principle, robust practical deployments for full, multi-step protocols remain difficult and comparatively scarce, with most demonstrations focusing on constrained unit operations rather than end-to-end recipes.}~\cite{refPINN} In parallel, explicit modelling of measurement noise and batch effects—for example via Gaussian-process discrepancy models—helps prevent overfitting to spurious experimental artefacts. Equally critical is rigorous uncertainty quantification—via deep ensembles, Monte-Carlo dropout or evidential networks—and meta-learning approaches such as model-agnostic meta-learning (MAML), which enable rapid adaptation to new reaction families with only a handful of additional experiments.~\cite{refAitken,refBelakaria}

\subsection{Physics-based simulators as priors and surrogates}
Beyond data-driven predictors, physics-based synthesis models—including kinetic Monte Carlo for surface reactions and nucleation, phase-field models for crystallisation, coarsening and porosity evolution, CALculation of PHAse Diagrams (CALPHAD)-informed diffusion models for multi-component solids, and computational fluid dynamics (CFD) for reactor-scale transport—provide mechanistic constraints and synthetic inductive bias. Today, these tools are computationally intensive, difficult to parameterise for novel chemistries, and often limited in predictive power for unknown, multi-component systems. Nevertheless, they are essential: they can (i) generate synthetic multi-fidelity datasets to pre-train protocol-aware predictors, (ii) act as differentiable or policy-evaluable environment models for BO and RL, (iii) provide mechanistic priors and constraints for $P\to X$ models (e.g., linking thermal histories to microstructural evolution), and (iv) identify kinetic traps and metastable pathways. Hybrid approaches with physics-informed neural networks (PINNs) and grey-box surrogates leverage partial differential equation structure for improved extrapolation and sample efficiency. \rev{These models should be judged not only by aggregate prediction error but also by whether they reproduce operando trajectories, transfer across chemistries or laboratories, and remain calibrated when $X$ is only partially observed.}
 
\subsection{Hybrid integration of structural knowledge with protocol-centric models}
Vast structural resources (e.g., Materials Project~\cite{ref6}, OQMD~\cite{refOQMD}) and DFT workflows must be \emph{central} to protocol-centric design, not peripheral. They contribute: (i) \textit{Phase-diagram and thermochemical constraints}: computed convex hulls, chemical potentials and phase boundaries restrict feasible target domains for $P\!\to\!X$ models; (ii) \textit{Metastable intermediates}: enumerated metastable phases suggest intermediate waypoints and targets for multi-step synthesis planning; (iii) \textit{Descriptor bootstrapping}: DFT-derived descriptors (formation energies, elastic moduli, redox potentials) augment $P$ representations when target structures are known; (iv) \textit{Route screening}: precursor compatibility and reaction driving forces from structure-property models prune infeasible branches in BO/RL planners; (v) \textit{Grey-box priors}: surrogate $X$ models warm-start on DFT/phase-field-consistent microstructural hypotheses that are refined by experimental data; and (vi) \textit{Generalisation}: structural embeddings help transfer knowledge across chemistries by conditioning protocol models on target-structure attributes.
\rev{In practice, however, hybrid integration is rarely plug-and-play: structural characterisation is often sparse, noisy, and multi-modal, and its primary value may be mechanistic understanding and hypothesis generation rather than immediate gains in purely predictive $P\!\to\!y$ accuracy.}
\rev{Accordingly, DFT databases and structure-derived embeddings should be treated as priors, filters or auxiliary descriptors rather than as substitutes for protocol-conditioned measurements, which remain the decisive ground truth for whether a synthesis route actually succeeds.}
 
\subsection{Inverse Design of Protocols}
 
In the synthesis protocol--property setting, inverse design corresponds to mapping target properties (and, where relevant, target structures) to one or more candidate protocols \(y^\star \!\to\! P^\star\) under experimental, safety and resource constraints. \rev{Table~\ref{tab:methods} should therefore be read as spanning three settings: direct $P\!\to\!y$ prediction, intermediate $P\!\to\!X$ modelling, and iterative or joint $P\!\to\!X\!\to\!y$ workflows.} Several complementary algorithmic families have emerged for this task:

\begin{itemize}
  \item \textbf{Bayesian optimisation} excels when the search space (continuous or discrete) is moderate in dimension and each experiment is costly. Extensions exist for multi-objective, multi-fidelity and constrained settings. Sequential model-based optimisation variants that allocate resources adaptively—such as Hyperband and Bayesian optimisation with HyperBand (BOHB)—or rely on Thompson sampling are popular choices for high-throughput SDLs.~\cite{refHyperband,refBOHB} Recent latent-space and trust-region formulations extend BO sampling efficiency to hundreds of design variables.~\cite{refLetham}
  \item \textbf{Reinforcement learning} treats synthesis planning as a sequential decision problem, ideal for multi-step routes and de~novo protocol construction. Reward design and sample inefficiency remain open challenges; model-based RL algorithms (e.g.~model-based policy optimisation (MBPO), Dreamer) mitigate sample inefficiency by learning differentiable environment models.~\cite{refZhou}
  \item \textbf{Generative models} (see Section~\ref{sec:generative}) can be conditioned on target properties to propose entirely new protocols. Ensuring chemical validity, experimental feasibility and safety of generated sequences is the foremost research hurdle.
  \item \rev{\textbf{Invertible and probabilistic inverse models:} conditional invertible neural networks and related flow-based approaches provide amortised approximate inversion with uncertainty estimates, complementing optimisation-based strategies when repeated inverse queries are required.}~\cite{refINN}
\end{itemize}
 
Rule-based symbolic retrosynthesis planners such as ASKCOS and AiZynthFinder provide chemistry-aware search primitives that can be combined with BO or RL for hybrid planning.~\cite{refASKCOS,refAiZynth} When differentiable process simulators are available, gradient-based optimisation of continuous protocol parameters is also feasible and has been demonstrated for photochemical flow reactors.~\cite{refBucci} For real-time robotic execution, optimisation algorithms must interface with job-shop scheduling or dead-time-aware planners and incorporate failure-recovery logic to maintain autonomous operation.~\cite{refScheduler}

As a concrete illustration, consider inverse design of a solid electrolyte synthesis protocol subject to practical constraints on furnace temperature, atmosphere and available unit operations. A protocol-aware optimiser can search over both composition and processing schedules to maximise ionic conductivity while simultaneously minimising process mass intensity and energy footprint, yielding Pareto fronts that trade off performance and sustainability. The algorithmic families summarised in Table~\ref{tab:methods} provide complementary routes to solving such constrained, multi-objective \(y^\star\!\to\!P^\star\) design problems.
 
 \begin{table}[ht]
   \centering\footnotesize
   \caption{AI/ML techniques for modelling and designing synthesis protocols.}
   \label{tab:methods}
   \setlength{\tabcolsep}{3pt}
   \begin{tabular}{>{\raggedright\arraybackslash}p{2.4cm} 
                   >{\raggedright\arraybackslash}p{3.1cm} 
                   >{\raggedright\arraybackslash}p{2.4cm} 
                   >{\raggedright\arraybackslash}p{3.5cm}
                   >{\raggedright\arraybackslash}p{3.8cm}}
     \hline
     Technique & Typical input & Output & Strengths & Key challenges \\
     \hline
     Transformers & Text, SMILES, actions & Property or protocol & Capture long-range context; pre-training & Need large data; hallucination \\
     GNNs & Reaction graphs & Property features & Respect chemical topology & Graph design, scalability \\
    Bayesian optimisation & Vector/tabular & Optimised parameters & Sample-efficient; uncertainty-aware & High-dimensional spaces; constraints; kernels for graphs/sequences \\
     Reinforcement learning & Action sequences & Policy (recipe) & Handles multi-step logic & Reward design; sample inefficiency \\
     Generative (VAE/GAN/Diff) & Latent vector + target property & Protocol & De-novo and conditional generation & Validity constraints; see Section~\ref{sec:generative} \\
     Symbolic planners & Target structure or property & Stepwise protocol & Chemically consistent; rule-based reasoning & Coverage limited by rule set; novelty constraints \\
    Simulation-based surrogates & Process simulator, physics model & Property or protocol & Mechanistic fidelity; gradient access; multi-fidelity priors & Simulator fidelity; parameter identifiability; compute expense \\
     Simulation-based optim. & Process simulator & Optimised protocol & in~silico risk-free evaluation; gradient access & Simulator fidelity; compute expense \\
     Causal ML & Protocol variables & Causal graph/effects & Mechanistic insight; extrapolation & Confounding; data requirements \\
     Foundation models & Text, graphs & Embedding / protocol & Leverage massive pre-training; data-efficient fine-tuning & Compute cost; domain adaptation \\
     \hline
   \end{tabular}
 \end{table}
 
\begin{figure}[ht]
   \centering
   \includegraphics[width=0.9\textwidth]{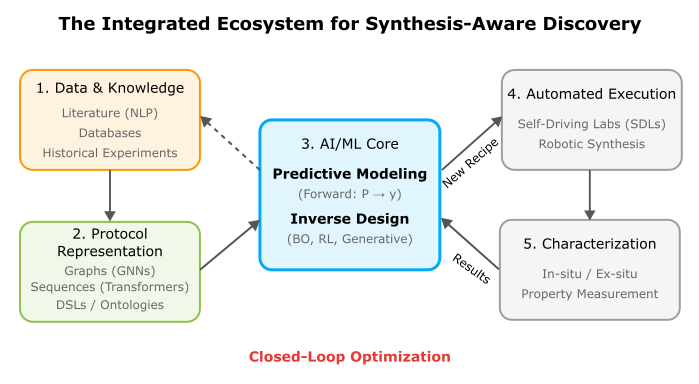}
\caption{Integrated ecosystem linking heterogeneous data sources, protocol representations, AI/ML core and automated execution/characterisation in a closed-loop discovery cycle. Characterisation streams feed both $P\!\to\!X$ and $X\!\to\!y$ models, enabling time-resolved updating during closed-loop operation.}
   \label{fig:ecosystem}
 \end{figure}

 Figure~\ref{fig:ecosystem} provides a high-level systems view of how these methodological components integrate into a cohesive synthesis-driven discovery workflow.

The methodological landscape surveyed in this section demonstrates that the AI/ML tools necessary for synthesis-first materials discovery are rapidly maturing. From protocol representations that capture the full complexity of inorganic synthesis to physics-informed models that bridge data-driven learning with mechanistic understanding, the technical infrastructure for the paradigm shift is falling into place. The convergence of these capabilities—coupled with the hybrid integration strategies that leverage decades of structure-centric knowledge—positions the field to move beyond the synthesizability gap toward truly predictive, experimentally grounded materials design. The following section examines how these tools are already delivering practical impact across diverse materials domains.
 
\section{From Vision to Reality: Applications and Early Indicators}
 \label{sec:cases}
 
The synthesis-centric paradigm is beginning to demonstrate practical impact across multiple materials domains. This section examines both early proof-of-concept demonstrations and mature applications.

\subsection{Early Proof-of-Concept Demonstrations}

Recent works provide compelling early indicators of protocol-aware AI's transformative potential. G{\'o}mez-Bombarelli \textit{et al.}~\cite{ref7} used variational autoencoders to generate multi-step polymer syntheses (organic analogue), while Tadanki \textit{et al.}~\cite{ref25} introduced \textsc{SynFormer}, a Transformer that outputs full reaction sequences. More directly relevant to inorganic materials, SDL platforms such as MINERVA seamlessly link protocol generation with robotic execution~\cite{refMinerva}, and autonomous flow platforms have demonstrated multistep quantum-dot synthesis in which protocol variables are repeatedly updated using in-line spectroscopic measurements that act as proxies for composition and particle state.~\cite{refAbolhasani} These successes, though nascent, confirm that embedding synthesis constraints can drastically improve experimental relevance compared to structure-only approaches.

\subsection{Mature Applications}

Below, we explicitly label application examples as (po) process/parameter optimisation or (dn) de~novo protocol design.

\subsubsection{Energy Storage and Conversion}

\textbf{Batteries (po-dn):} Closed-loop optimisation of fast-charging protocols with ML improves performance and mitigates degradation in Li-ion cells.~\cite{ref67} In continuous-flow settings, combining BO with mechanistic surrogate models reduces experiments and navigates multi-objective trade-offs.~\cite{ref70} Here, the protocol variables themselves are the design space, while electrochemical response and process-state measurements provide the feedback needed to iteratively refine the search; this is often a direct $P\!\to\!y$ setting, but can be enriched toward $P\!\to\!X\!\to\!y$ when degradation state or microstructural proxies are explicitly modelled.

\noindent\textbf{Photovoltaics (po):} Automated spin-coating coupled to BO has been used to co-optimise precursor composition and processing conditions for perovskite solar cells, including antisolvent handling, spin speed and annealing temperature.~\cite{refPV} This example makes the $P\!\to\!X\!\to\!y$ logic explicit: the protocol governs crystallisation and film formation, time-resolved fluorescence and related film-quality measurements provide structure-sensitive proxies of $X$, and power-conversion efficiency supplies the downstream property target.

\subsubsection{Functional Electronic Materials}

\textbf{Semiconductors and quantum dots (po):} Machine-learning-assisted real-time feedback control has been used to steer InAs/GaAs quantum-dot growth by coupling growth conditions to in situ reflection high-energy electron diffraction (RHEED) videos, which act as structure-sensitive observations of the evolving surface and quantum-dot density.~\cite{ref74} This is precisely the regime in which explicit treatment of $X$ is integral rather than optional: the controller does not merely learn $P\!\to\!y$, but adjusts the protocol in response to measured proxies of intermediate structural state before the final material specification is achieved.

\subsubsection{Catalytic and Surface-Active Materials}

\textbf{Catalysis (po):} Active learning coupled to high-throughput experimentation accelerates discovery and optimisation of electrocatalysts for CO$_2$ reduction and hydrogen evolution.~\cite{ref78} Interpretable ML (e.g., Shapley additive explanations, SHAP) elucidates process-structure-property relationships to guide catalyst optimisation.~\cite{ref15} These examples highlight how protocol-aware approaches can reveal process-structure-performance correlations invisible to equilibrium structure-only models, especially when paired with automated electrochemical workstations and inline product analysis.

\subsubsection{Advanced Functional Materials}

\textbf{Magnetic materials (po):} While some examples remain structure-centric (ML-guided DFT screening for rare-earth-free magnet candidates~\cite{ref83}), AI methods are beginning to accelerate candidate identification in 2D magnetic materials where processing conditions critically affect magnetic ordering.~\cite{ref3} Protocol-centric approaches that couple processing schedules with magnetometry and transport measurements offer a natural path to closed-loop optimisation in this space.

\noindent\textbf{Metal-organic frameworks (MOFs) (po, emerging dn):} Closed-loop and BO-driven strategies demonstrated in related solvothermal and flow systems provide a template for MOF crystallisation optimisation.~\cite{refBucci} MOFs represent an ideal testbed for synthesis-centric approaches because crystallisation outcomes depend sensitively on temperature ramps, pH evolution, and nucleation kinetics, all of which can be monitored via inline scattering or spectroscopy.

\noindent\textbf{Biomaterials and drug delivery (po):} The MINERVA SDL automates synthesis and inline characterisation of nanomaterials for delivery applications,~\cite{refMinerva} while ML-guided photoinduced electron/energy transfer-reversible addition-fragmentation chain-transfer (PET-RAFT) polymerisation in a robotic workstation rapidly generates polymer libraries for bio-interface screening.~\cite{ref60} Though focusing on organic polymers, these examples demonstrate the power of protocol-centric automation, where robotic execution logs, inline assays and high-throughput screening data jointly fuel protocol optimisation.

\subsection{Emerging Patterns and Lessons Learned}

Several key patterns emerge from these applications: (i) \textit{Flow and continuous processing} environments are particularly amenable to protocol-centric optimisation because they enable real-time parameter adjustment; (ii) \textit{Multi-objective optimisation} (performance, sustainability, cost) is naturally handled when economic and environmental factors are embedded in protocol representations; (iii) \textit{Closed-loop operation} with inline characterisation provides the rapid feedback necessary for effective learning; (iv) \textit{Interpretable models} that reveal process-structure-property relationships offer scientific insights beyond pure optimisation.

The most successful examples share a common architecture: protocol representations that capture the essential physics, predictive models trained on high-quality experimental data, and optimisation algorithms that can navigate multi-dimensional parameter spaces efficiently. Critically, the highest-impact applications focus on problems where processing conditions strongly influence final properties—precisely where structure-only approaches fail.

These case studies demonstrate that protocol-aware AI is transitioning from proof-of-concept to practical impact across diverse materials domains. The synthesis-centric paradigm is not merely a theoretical alternative but a demonstrated pathway to bridging the synthesizability gap and accelerating materials discovery.
 
 
\section{Challenges and Future Directions}
\label{sec:challenges}
 
Adoption of the synthesis protocol-property paradigm faces several non-trivial hurdles, which we categorize by priority and outline below. Table~\ref{tab:challenges} summarises challenges and prospective research directions.

\subsection{Critical Near-Term Challenges}

\textbf{Data scarcity, quality and accessibility.} This represents the most immediate barrier to widespread adoption. Existing literature is dominated by unstructured prose, publication bias toward positive results, missing metadata, and proprietary restrictions. Furthermore, data pulled from text-mined literature are heterogeneous and biased toward successful experiments, whereas self-driving laboratories yield highly structured but domain-specific datasets. We recognise that, in contrast to million-entry structure databases such as Materials Project~\cite{ref6} or OQMD~\cite{refOQMD}, protocol-property corpora remain smaller and noisier, but they are expanding rapidly via automated NLP extraction and SDL campaigns. \rev{Robust modelling under these conditions requires more than simple data accumulation: it demands deliberate capture of failed, null and low-yield experiments, chemistry-aware reweighting or stratified sampling, and evaluation protocols that distinguish interpolation within one laboratory from genuine transfer across platforms. The value of failed experiments for materials ML has already been demonstrated in adjacent settings, and protocol-centric discovery will benefit from adopting the same lesson early.}~\cite{refRaccuglia}

\rev{\noindent\textbf{Why structure-centric ML has dominated.}
The community's historical fixation on structure is not accidental: crystal structures are comparatively well standardised; structure-derived labels (energies, band gaps, elastic constants) can be generated at scale with mature simulation workflows; and many property predictors are naturally formulated as structure$\to$property mappings. By contrast, protocol data are heterogeneous, platform-dependent and often underspecified, making protocol-centred AI both harder and more brittle. Recognising these structural reasons is important for designing realistic roadmaps that prioritise standardisation, provenance and robust execution.}

\rev{\noindent\textbf{Provenance and event-driven architectures.}
Protocol interoperability across laboratories requires more than shared ontologies: it requires reliable capture of \emph{execution provenance} (what actually happened, with what deviations), and durable links between synthesis events and downstream characterisation. Event-sourced and streaming architectures adapted from distributed systems have been proposed for materials provenance and experiment control (e.g., ESAMP and MDML), but remain complex to implement and have not yet seen widespread community adoption.}~\cite{refESAMP,refMDML}

\noindent\textbf{Heterogeneity and multi-fidelity learning.} Text-mined protocols are noisy and incomplete, whereas SDL logs are structured but platform-specific. We advocate hierarchical Bayesian or Gaussian-process multi-fidelity modelling to fuse literature-scale low-fidelity data with high-fidelity SDL measurements, coupled to explicit measurement-error models and batch-effect correction. Standardised protocol normalisation (units, timing semantics, vessel context) and schema versioning are essential to prevent dataset drift and enable reproducible training. Ambiguous NLP parses benefit from human-in-the-loop curation and active error-correction workflows. \textit{Primary challenge:} Achieving robust transfer across domains— from narrow SDL regimes to literature-scale chemistry—requires principled domain adaptation and transfer learning, with careful handling of covariate and label shifts.
 
\noindent\textbf{Standardisation of protocol representation.} The diversity of text snippets, SMILES strings, reaction graphs and action sequences hampers model interoperability. Consensus ontologies and parsers capable of round-tripping between formats will be crucial to avoid siloed datasets.~\cite{ref8}
 
\subsection{Medium-Term Technical Challenges}
 
\textbf{Model generalisability and interpretability.} Deep models trained on narrow chemical domains often fail on unseen reactions. Incorporating physical priors, providing uncertainty quantification and employing interpretable architectures (e.g., latent-space attention visualisation, SHAP) will build trust and insight.~\cite{ref15}
 
\noindent\textbf{Integration with automated experimentation.} SDLs promise high-throughput, reproducible data, yet real-world robotic platforms still incur nontrivial failure rates (10--30\,\%), require human-in-the-loop oversight, and often lack chemist-friendly interfaces. In the medium term we therefore envision \emph{human--AI collaboration} as the dominant operating mode: AI assists chemists with planning, monitoring, and failure recovery rather than operating fully lights-out. Beyond algorithmic integration, there is an \emph{executability gap}: mapping abstract protocols to reliable robotic execution under hardware variability (dispense precision, dead volumes, thermal lags, calibration drift). For inorganic workflows, solids handling (grinding/milling, weighing powders), pellet pressing, high-temperature furnace loading/unloading, atmosphere control and thermal profile verification remain difficult to automate robustly compared to liquid handling. Practical adoption will require job-shop scheduling with dead-time awareness, automatic calibration and verification routines, and fail-safe recovery policies; these engineering controls should co-evolve with protocol-aware AI.
 
\noindent\textbf{Scalability and computational cost.} Training large generative models and running BO across high-dimensional spaces demand significant compute. Protocol-centric models—often featuring multimodal embeddings, simulator calls, or experiment-in-the-loop optimisation—can incur higher training and deployment costs than structure-only VAEs or diffusion networks; however, multi-fidelity, surrogate-model, and lightweight graph-based strategies can mitigate these burdens. \rev{The comparison is also subtle: structure-first models may be cheaper per inference, but protocol-aware systems can lower end-to-end discovery cost if they rapidly avoid uninformative or failed experiments. For this reason, the relevant metric is not model floating point operations (FLOPs) in isolation but scientific yield per unit of combined laboratory and compute budget.} We further note that scale-up challenges (heat/mass transfer, economic amortisation of equipment and reagents) must be considered when evaluating protocol viability.
 
\noindent\textbf{Benchmarking and standards.} Objective comparison of new algorithms demands curated benchmark suites and clear evaluation protocols. Although reaction-informatics and autonomous-experimentation communities have begun to assemble shared datasets and benchmark tasks, coverage across inorganic synthesis and protocol-aware inverse design remains sparse. The more immediate priority is therefore to define benchmark tasks, transfer settings, calibration metrics and reporting conventions that reflect real experimental constraints rather than purely computational scores.
 
\subsection{Future Research Directions}

The challenges outlined above point to several high-priority research directions that will determine the success of the synthesis-centric paradigm: How best to encode multi-step, multi-phase protocols for universal learning? Can generative AI be constrained to obey green chemistry metrics during protocol generation? How do we merge heterogeneous data streams (text, graphs, SDL logs) into unified training corpora that preserve the strengths of each modality? What active-learning strategies can navigate highly non-convex, constrained synthesis spaces most effectively? And perhaps most critically, how can we develop transferable models that generalise across chemical domains while maintaining the mechanistic insights that make protocol-centric approaches scientifically valuable?

\rev{\noindent\textbf{Can there be a foundation model for protocol space?}
Large-capacity models spanning broad regions of molecular and materials structure space naturally motivate analogous questions for synthesis.~\cite{refMACE} We consider protocol-space foundation models plausible, but harder to universalise than structure models because protocols depend on platform-specific action vocabularies, incomplete provenance, temporal control logic and the coupling of textual, sensor and hardware data. In our view, the most realistic route is not a single monolithic model trained on "all protocols", but interoperable foundation models grounded in execution logs, characterisation streams and explicit ontologies that permit transfer across platforms while preserving uncertainty about out-of-distribution actions.}~\cite{refXDL,refESAMP,refMDML}

\rev{\noindent\textbf{Can synthesis be simulated from "structure-first" models?}
We view protocol-first and structure-first approaches as complementary. It is plausible that future multi-scale simulation—enabled by major advances in algorithms, parameterisation and compute—could capture aspects of nucleation, growth and processing with increasing fidelity. However, present-day atomistic and continuum tools rarely provide end-to-end predictive control over synthesis outcomes for complex inorganic systems. Protocol-centred modelling therefore offers a pragmatic route to progress in the near and medium term, while leaving open the long-term possibility of deeper first-principles synthesis simulation.}

\noindent\textbf{Concrete community to-do list.}
We highlight three near-term priorities for standards and benchmarking: (i) convergence toward interoperable protocol ontologies and grammars (e.g., XDL, PAML, AnIML) with robust round-tripping between text, graph and action-sequence formats; (ii) SDL logging standards that capture fault-tolerant execution traces, including failures and recovery actions, to support reliable $P\!\to\!X\!\to\!y$ model training; and (iii) open, community-maintained benchmark suites for protocol-aware modelling and inverse design that include negative and null results, not only optimised successes.

Addressing these challenges will require coordinated efforts across the materials science, machine learning, and automation communities. Success will depend not only on algorithmic advances but also on community-wide adoption of data standards, collaborative benchmark development, and sustained investment in the infrastructure necessary to support synthesis-aware discovery at scale.
 
 \begin{table}[ht]
   \centering\footnotesize
   \caption{Summary of challenges and prospective research directions.}
   \label{tab:challenges}
   \renewcommand{\arraystretch}{1.1}
   \begin{tabular}{>{\raggedright\arraybackslash}p{3.3cm} >{\raggedright\arraybackslash}p{4.5cm} >{\raggedright\arraybackslash}p{5.5cm}}
     \hline
     \textbf{Challenge area} & \textbf{Specific issues} & \textbf{Proposed directions} \\
     \hline
     Data availability & Unstructured text; few negative results & findable, accessible, interoperable, and reusable (FAIR) protocols; NLP extraction; SDL data generation \\
     Representation & No universal format & Community ontologies; conversion tools; benchmarking \\
     Generalisability & Domain shift, black-box models & Physics-informed nets; uncertainty quantification; interpretable ML \\
     Automation integration & Hardware/software incompatibility & Modular lab OS; human-in-the-loop; fault recovery \\
     Data fusion \& transferability & SDL narrow vs literature broad; domain shift & Multi-fidelity GP/Bayes; transfer learning; domain adaptation; meta-learning \\
     Computational cost & Large models; expensive BO loops & Surrogate models; multi-fidelity; efficient architectures \\
     \hline
   \end{tabular}
 \end{table}
 
 
\section{Conclusion: Toward a Synthesis-First Future}
\label{sec:conclusion}
 
The prevailing structure-property paradigm has delivered profound scientific insights and enabled remarkable computational advances, but it falters when predictions confront experimental reality. The persistent synthesizability gap—where fewer than 2\% of computationally predicted phases are ever realised—represents not merely a technical hurdle but a fundamental architectural limitation of structure-centric approaches. By elevating synthesis protocols to first-class citizens in modelling and inverse design, the synthesis-first paradigm embeds feasibility constraints from the outset, offering a direct pathway to bridge this divide.

The evidence presented in Sections~\ref{sec:generative}--\ref{sec:cases} demonstrates that this paradigm shift is both necessary and achievable. Early successes—from protocol-optimised battery electrolytes to self-driving quantum dot synthesis—confirm that embedding experimental constraints dramatically improves hit rates compared to structure-only approaches. The methodological infrastructure is rapidly maturing: protocol representations that capture inorganic synthesis complexity, physics-informed models that bridge data-driven learning with mechanistic understanding, and hybrid integration strategies that leverage decades of structure-centric knowledge. These advances collectively point toward an accelerated, more reliable route to functional materials.

\textbf{A vision for 2035.} We envision a materials discovery ecosystem where synthesis protocols are the primary design variables, increasingly integrated with autonomous experimentation and real-time characterisation. In this future, materials scientists design complete, executable recipes for target properties more routinely than today, with AI systems that not only predict thermodynamic stability but also suggest feasible routes that account for kinetics, sustainability and cost. Self-driving laboratories operate as extensions of human creativity, exploring protocol spaces guided by interpretable models that reveal process-structure-property relationships. We expect the synthesizability gap to narrow substantially as these elements mature and interoperate.

\rev{\textbf{A call to the community.} Realising this vision requires coordinated action across the materials science, machine learning, and automation communities. In particular, progress will be limited if advances remain purely algorithmic: robust, scalable experimentation platforms and interoperable protocol execution are equally central. We therefore call for: (i) sustained investment in reliable, high-throughput self-driving laboratory infrastructure and fault-tolerant operation; (ii) interoperable protocol representations, execution layers and provenance/logging standards that enable workflows to transfer across labs; (iii) shared protocol-property databases and standardised representations; (iv) community benchmarks that prioritise experimental validation over computational metrics; (v) integration of synthesis-aware curricula in materials science education; (vi) collaborative development of open-source tools that democratise access to synthesis-centric AI; and (vii) policy frameworks that incentivise data sharing while protecting intellectual property. The transition from structure-centric to synthesis-first materials discovery represents not just a technological shift but a cultural transformation—one that will require the collective commitment of our community to achieve.}

By embracing protocols as design objects and embedding experimental reality into our computational frameworks, we can transform materials science from a discipline of discovery to one of design—one executable recipe at a time.

\bigskip

\section*{Conflicts of interest}
The author declares no conflicts of interest.

\section*{Data availability}
No new datasets, software, or code were generated or analysed in this perspective. All information is contained within the article and its references. If future code or data deposits are created (e.g., figure source files or synthetic benchmark corpora), the corresponding repository links and DOIs will be provided in the final submission.

\bigskip

\bibliographystyle{unsrt}
\bibliography{references_JPM}

\end{document}